\newfont{\mathbb}{bbold12}
\newfont{\Bbf}{bbold8}
\newcommand{\intR}{\mbox{\Bbf R}} 
\newcommand{\R}{\mathbb R}    
\newcommand{\C}{\mathbb C}    

\newcommand{\Pk}{\mathbb P}

\newcommand{\CP}{\C\Pk^{1}}
\newcommand{\CPn}{\C\Pk^{n}}

\newcommand{\One}{I}
\newcommand{\Two}{I\!\!I}
\documentclass[12pt]{JHEP}
\usepackage{epsfig,amsfonts,mathrsfs}
\title{Comments on Noncommutative Sigma Models}
\author{Jeff Murugan \\
	Department of Mathematics and Applied Mathematics\\
	University of Cape Town\\ Private Bag, Rondebosch 7700\\ 
	South Africa\\
	\email{jeff@hbar.mth.uct.ac.za}}
\author{Rory Adams \\ Department of Physics\\
        University of Cape Town\\ Private Bag, Rondebosch 7700\\ 
	South Africa\\ 
        \email{rory@casimir.phy.uct.ac.za}}	

\abstract{
We review the derivation of a noncommutative version of the 
nonlinear sigma model on $\CPn$ and it's soliton
solutions for finite $\theta$ emphasizing the similarities it bears to the GMS scalar field theory. It is also shown that unlike the scalar theory, some care needs to be taken in defining the topological charge of BPS solitons of the theory due to nonvanishing surface terms in the energy functional. Finally it is 
shown that, like its commutative analogue, the noncommutative $\CPn$-model also
exhibits a non-BPS sector. Unlike the commutative case however, there are
some surprises in the noncommutative case that merit further study.    
} 
  
\begin{document}

\section{Introduction}
\label{Introduction}
\noindent
Nonlinear sigma models on K\"ahler and hyper-K\"ahler target spaces are arguably
some of the most important test beds of ideas that invariably find their way into the
more daunting arena of physical gauge theories in four dimensional spacetimes.
Certainly one of the most favored of such theories is the $d=2$ sigma model with 
target space $\CPn$ - the $n$-dimensional complex projective space. 
Like the $d=4$ self-dual Yang-Mills 
theory it too exhibits asymptotic freedom, conformal invariance
and a rich solitonic sector.\\

\noindent
A large class of the soliton solutions of the $\CPn$ model are the finite energy lump-like solutions that correspond to holomorphic functions on the two-dimensional base space. The lumps saturate a BPS bound on the $\CPn$ energy functional and are consequently stabilized by some finite topological charge. Although these are by far the most well studied, they are by no means the only solitonic solutions exhibited by the $\CPn$ sigma model. It has been known for some time that certain bound states of such BPS lumps also solve the sigma model equations \cite{DinZak1}. These are, however, not solutions of any first order BPS equations and consequently lack the stability properties of the lumps. Nevertheless, there exists a B\"acklund-like solution generating technique for generating general (non-BPS) $\CPn$ solutions from a given holomorphic BPS soliton \cite{DinZak2,Sasaki}. Yet even in the light of such remarkable similarities between the $\CPn$sigma model and four-dimensional gauge theories, some differences are quite stark. Chief among these are the lack of a more complete understanding of the soliton moduli space and the absence of a general construction technique like the ADHM method for the $\CPn$ model. One avenue toward a better understanding of the dynamics of the $\CPn$ lumps (as encoded in the moduli space) lies in the deformation of the base space on which the lumps move.\\ 

\noindent
Ever since the realization that the low energy effective theory of D-branes in a $B$ field background \cite{SeibergWitten,Witten} is a noncommutative field theory, the deformation of choice has become that of the algebra of smooth functions over the base. This yields a noncommutative $\CPn$ sigma model whose basic solitonic excitations have by now been well documented \cite{LeeLeeYang}. In particular, the moduli space metric was explicitly computed for the $1-$ and $2-$soliton solutions and shown to be nonsingular and K\"ahler in both cases \cite{FurutaInami}. Moreover, in \cite{FurutaInami2} it was shown that, in stark contrast to the commutative case, the noncommutative $\CP$ sigma model contains a non-BPS sector that is closely tied to the scalar solitons of the GMS field theory \cite{GMS}. The existence of these new non-BPS excitations of the $\CP$ model (and, more generally in the $\CPn$ model) is certainly intriguing. If nothing else, it is a reminder of the fact that the volume of the solution space of the noncommutative theory is significantly larger than the corresponding commutative one. An interesting question then, is whether the known soliton generating technique of \cite{DinZak2,Sasaki} probe this sector of the solution space of the noncommutative $\CPn$ sigma model. As will be demonstrated, this technique is, perhaps surprisingly, deficient in the noncommutative model.\\    

\noindent
Despite (or perhaps because of) their remarkable simplicity, GMS solitons have had a huge impact on recent literature (see \cite{Harvey1} for a recent review). In particular, it was shown in \cite{Bonora1,Bonora2} that the algebraic structure of a family of solitonic solutions of the vacuum string field theory equations
\begin{eqnarray}
  \Psi_{m}* \Psi_{m} = \Psi_{m}
  \label{VSFT}
\end{eqnarray}
\noindent
is exactly isomorphic to the corresponding one for the GMS solitons of a noncommutative pure scalar field theory. Exploiting this isomorphism leads one to the interpretation of such noncommutative solitons as relics of $D23$-branes in the low energy limit. If, as in \cite{FurutaInami2} (and later on in this paper), solitonic excitations of the $\CPn$ sigma model exist that can be built up of bound states of scalar solitons, it seems natural to ask whether the noncommutative sigma model solitons may have some interpretation as $D$-brane configurations also.\\
         
\noindent
The organization of this paper is as follows: After a brief description of the
$\CPn$ sigma model and its (commutative) instanton solutions, we proceed to a
review of the corresponding noncommutative instantons. While the results in this section are themselves not new, the {\it formulation} of the noncommutative sigma model is. By focusing on the formal similarity between the sigma model equations and that of the noncommutative scalar field theory, the BPS bound on the energy functional is rewritten to emphasize the subtlties encountered in defining topological charges of noncommutative objects. This section will establish all
the necessary formalism required for the main result of this work: the
construction of non-BPS solitons of the noncommutative $\CPn$ sigma
model{\footnote{As this work was nearing completion we became aware of
the work of Foda {\it et. al.} \cite{FodaJackJones} whose results have
significant overlap with our own. The emphasis in \cite{FodaJackJones} is
largely on demonstrating that many of the the known results for the 
construction of general solitonic solutions to the $\CPn$ sigma model are 
equally applicable in the
noncommutative case. The point of our work however, is to highlight the
similarities in the description of the BPS and non-BPS solitons of the
noncommutative sigma model to that of the scalar GMS solitons hopefully paving the
way for further study into the possible embedding of the former into a stringy framework \cite{Townsend}}}. After the construction of several explicit non-BPS solitons for both the $\CP$ and $\C\Pk^{2}$ sigma models, the following section is devoted to the comparison of the
BPS solitons constructed as holomorphic curves on $\CPn$ 
and those obtained from bound states of GMS scalar solitons. 
  
\section{The noncommutative $\CPn$ instanton}
\label{Sec1}
\noindent
 By way of establishing notation and 
 some of the conventions, to be followed
 for the remainder of this paper, we begin by reviewing the construction of
 the noncommutative soliton solutions of the nonlinear sigma model on a
 $\CPn$ target space. This section follows closely the recent work of 
 Lee, Lee and Yang{\cite{LeeLeeYang}}.\\

\subsection{Notation} 
\noindent
 In studying noncommutative nonlinear sigma models we will, for the most part, 
 be interested in maps $u:\,\R^{2}_{\theta}\times\R\,\rightarrow M$ with the target, 
 $M$ 
 a K\"ahler (or hyper-K\"ahler) manifold. The sigma model field $u$ takes values 
 in the $\theta$-deformed algebra of functions over $\R^{2}_{\theta}$, ${\cal
 A}_{\theta}$, whose elements satisfy
\begin{eqnarray}
 f\star g(x) =
 e^{\frac{i}{2}\theta^{ij}\partial_{i}\partial'_{j}}f(x)g(x')|_{x=x'}\,,
 \label{Moyal}
\end{eqnarray}
\noindent
 where $\theta^{ij} = \theta\epsilon^{ij}$ is a nondegenerate, antisymmetric
 constant matrix and $\theta$ is a positive
 deformation (noncommutativity) parameter of dimension $(mass)^2$. 
 Consequently, coordinates on the noncommutative plane $\R^{2}_{\theta}$ 
 satisfy the Heisenberg algebra 
 $[x^{1},x^{2}] := x^{1}\star x^{2} - x^{2}\star x^{1} = i\theta$. 
 Written in terms of the complex coordinates $z:=(x^1+ix^2)/\sqrt{2}$ and
 $\bar{z}:=(x^1-ix^2)/\sqrt{2}$ on $\R^{2}_{\theta}$ the commutator
 becomes $[z,\bar{z}] = \theta$ which, up to a rescaling
 is nothing but the algebra of annihilation and creation operators for the
 simple harmonic oscillator. 
 By use of the Weyl transform \cite{Harvey1}, we associate to a function on the
 noncommutative space an operator acting on an auxiliary Hilbert space ${\cal H}
 = L^{2}(\R)$. In a basis of simple harmonic oscillator eigenstates ${\mathcal{H}} =
 \bigoplus_{n}\C|n\rangle$. The vacuum 
 $|0\rangle$ is defined, as usual, by the action of the annihilation 
 operator $\widehat{z}$ on it as $\widehat{z}|0\rangle = 0$. Further, we have 
\begin{eqnarray}
   \widehat{z}|n\rangle &=& \sqrt{\theta n}|n-1\rangle\,,\\
   \widehat{\bar{z}}|n\rangle &=& \sqrt{\theta(n+1)}|n+1\rangle\,.
\end{eqnarray}  
\noindent
 This association of functions on the noncommutative space and operators in the
 Hilbert space is particularly useful in treating differentiation and
 integration on the noncommutative plane. Under the Weyl map 
 the operations of differentiating and integrating functions 
 over $\R^{2}_{\theta}$ transform to 
\begin{eqnarray}
 \partial_{i} &\rightarrow& 
 {\frac{i}{\theta}}\epsilon_{ij}[\widehat{x^{j}},\cdot\,]\,,\\
 \int_{\intR^{2}_{\theta}}d^{2}x\,f(x^{i}) &\rightarrow& 2\pi\theta\,\,
 {\rm Tr}_{\cal H}\,\widehat{O_{f}}(\widehat{x^{i}}) = 
 2\pi\theta\sum_{n\geq0}\,\langle n|\widehat{O_{f}}|n\rangle\,. 
 \label{Weylcorr2}
\end{eqnarray}    
\noindent
 In particular, tracing over the Hilbert space preserves the translational
 symmetry of the noncommutative plane.

\subsection{The commutative $\CPn$ sigma model}
\noindent
In this section we collect some well known results on classical nonlinear sigma
models on K\"ahler target spaces \cite{Townsend,Ward85,Ioannidou} that will prove
useful in what follows. If $X: \R^{(1,2)}\rightarrow M$ is a map from $(2+1)$-dimensional
Minkowski spacetime with standard metric $\eta_{\mu\nu} = diag(-1,+1,+1)$ 
to a K\"ahler target
manifold with Riemannian metric $g_{IJ}$ then the action for the nonlinear sigma
model is
\begin{eqnarray}
 S = \frac{1}{2}\int_{\intR^{(1,2)}}\!\!\!d^{3}x\,\,
 \eta^{\mu\nu}\partial_{\mu}X^{I}\partial_{\nu}X^{J}g_{IJ}\,.
 \label{SMAction1}
\end{eqnarray} 
\noindent
The K\"ahler property of the target manifold means that there exists a covariantly
constant real $(1,1)$-tensor field (the almost complex structure) $J$ 
satisfying $J^{I}_{K}J^{K}_{L} 
= -\delta^{I}_{L}$ and a closed real two form (the K\"ahler form) $\Omega =
\frac{1}{2}J_{IK}dX^{I}\wedge dX^{K}$. In terms of the almost complex structure and
the K\"ahler form, the energy of a static field configuration may be rearranged to
give
\begin{eqnarray}
 E = \frac{1}{4}\int_{\intR^{2}}\!\!d^{2}x\,(\partial_{i}X^{I} 
 \pm \epsilon^{j}_{i}J^{I}_{K}\partial_{j}X^{K})^{2} \mp
 \underbrace{\frac{1}{2}\int_{\intR^{2}}\!\!d^{2}x\,\Omega_{IK}\epsilon^{lm}\partial_{l}X^{I}
 \partial_{m}X^{K}}_{2\pi Q}\,.
 \label{BogomolnyiBound}
\end{eqnarray} 
\noindent 
The second term (the topological charge) is just the integral over $\R^{2}$
of the pullback of the K\"ahler form and is a topological invariant as a result of the
fact that $\Omega$ is a closed form. This gives the familiar bound on the energy
$E\geq 2\pi|Q|$. The energy bound is saturated by configurations that 
satisfy the BPS equations
\begin{eqnarray}
 \partial_{i}X^{I} 
 \pm \epsilon^{j}_{i}J^{I}_{K}\partial_{j}X^{K} = 0\,.
 \label{KahlerBPS}
\end{eqnarray}
\noindent
Since these are the just the Cauchy-Riemann equations, such configurations are
nothing but holomorphic curves on the K\"ahler manifold $M$. Now fix $M$ to be the
$n$-dimensional complex projective space $\CPn = \C^{n+1}/\C^{*}$. In terms of the
sigma model fields $X^{I}(z,\bar{z})\,,\,I=1,\ldots,n$ (the inhomogeneous coordinates on 
$\CPn$) the standard Fubini-Study metric is given by
\begin{eqnarray}
 ds^{2} = 4\frac{\delta_{IJ}(1+\overline{X}_{K}X_{K})-\overline{X}_{I}X_{J}}
 {(1+\overline{X}_{K}X_{K})^{2}}dX^{I}d\overline{X}^{J}\,.
\end{eqnarray}   
\noindent
The sigma model action is most conveniently formulated in terms of the $\CPn$ 
homogeneous coordinates $U = (u_{1},\ldots,u_{n+1})\sim (\lambda u_{1},\ldots,\lambda
u_{n+1})$ where $\lambda\in \C^{*}$ is a nonzero complex number. Defining
$DU:=dU+iUA$, this is given by
\begin{eqnarray}
  S = \int_{\intR^{(1,2)}}\,d^3x\,\eta^{\mu\nu}(D_{\mu}U)^{\dagger}D_{\nu}U \,,
  \label{GIClassicalPnAction}
\end{eqnarray}
\noindent
subject to the constraint $U^{\dagger}U - 1 = 0$. A few points should be immediately
apparent from this formulation; the first being the invariance of the action 
under global $SU(n+1)$
transformations of the sigma model fields $u_{I}\rightarrow e^{i\alpha}u_{I}$. This is
merely a reflection of the equivalence relation defining $\CPn$. The second being the
fact that the `gauge field' is an auxiliary one, completely determined by the sigma
model fields $A = iU^{\dagger}dU$. The corresponding equations of motion
written in terms of the (homogeneous) sigma model fields are given by
\begin{eqnarray}
 D_{\mu}D^{\mu}U + (D_{\mu}U)^{\dagger}(D^{\mu}U)U = 0\,.
\end{eqnarray}  
\noindent
Once again the static energy is bounded by a topological charge $E\geq 2\pi|Q|$ where
now
\begin{eqnarray}
 Q =
 \frac{i}{2\pi}\int_{\intR^{2}}\!\!d^{2}x\,\epsilon^{ij}(D_{i}U)^{\dagger}(D_{j}U)\,.
\end{eqnarray}
\noindent
Reparameterising the sigma model field $U = W/\sqrt{W^{\dagger}W}$ where $W$ is an
$(n+1)$-vector, the energy bound is saturated when the first order BPS equations
$\partial_{\bar{z}}W = 0$ or $\partial_{z}W = 0$ are satisfied. These are the
instanton and anti-instanton solutions of the $\CPn$ sigma model, constructed by 
taking $W$ to be a rational function of $z$ and $\bar{z}$ respectively. The topological
charge of the soliton is counted as the highest degree of the rational function
components of $W$. Before discussing noncommutative generalizations it is worth 
noting that the $\CPn$ sigma model may be formulated
completely in terms of the Hermitian projector $P = W(W^{\dagger}W)^{-1}W^{\dagger}$
in terms of which the action is given by
\begin{eqnarray}
  S = \frac{1}{2}\int_{\intR^{(1,2)}}d^3x\,\,{\rm tr}
  \,\,\eta^{\mu\nu}{\Bigl(}\partial_{\mu}P\partial_{\nu}P{\Bigr)}\,.
  \label{ProjAction}
\end{eqnarray}   
\noindent
The `trace' in the integrand is the usual matrix trace operation and the unitary
constraint on the sigma model fields $U(z,\bar{z})$ is reflected in $P^2 = P$. This
formulation will prove particularly useful in the construction of non-BPS solitons
later.

\subsection{The noncommutative $\CPn$ sigma model and its BPS solutions}
The transition to a noncommutative $\CPn$ sigma model is made, following the standard
prescription, by replacing all products occurring in the above formulae with Moyal
$\star$-products and subsequently by replacing all noncommutative functions with the
associated operators on ${\cal H}$. As such, the sigma model action 
(\ref{GIClassicalPnAction}) becomes
\begin{eqnarray}
 S_{\theta} = \frac{2\pi}{\theta}{\rm Tr}_{\cal H}
 {\Bigl(}\delta_{ij}[\widehat{x^{i}},\widehat{U}^{\dagger}](1-P)
 [\widehat{U},\widehat{{x}^{j}}]{\Bigr)}\,.
 \label{OperatorPnAction}
\end{eqnarray}
\noindent
The unitarity condition on the commutative sigma model fields $U^{\dagger}U = 1$
becomes an isometry $\widehat{U}^{\dagger}\widehat{U} = 1$ on ${\cal H}$ 
(see \cite{Harvey1} for more details). In deriving (\ref{OperatorPnAction}) 
use was made of the identity
$D_{i}U\rightarrow \frac{i}{\theta}\epsilon_{ij}(1-P)[\widehat{x^{j}},\widehat{U}]$
and $P$ is the Hermitian projector as defined above. As in the commutative case, the
static action may be rewritten in completely in terms of $P$ as
\begin{eqnarray}
 S_{\theta} = 2\pi {\rm Tr}_{\cal H}{\rm tr}{\Bigl(}
 [P,\widehat{a}^{\dagger}][\widehat{a},P]
 {\Bigr)}\,,
 \label{OperatorProjAction} 
\end{eqnarray}
\noindent
after a further rescaling of the coordinates on $\R_{2}^{\theta}$ as
$\widehat{z}\rightarrow\sqrt{\theta}\widehat{a}$ and
$\widehat{\bar{z}}\rightarrow\sqrt{\theta}\widehat{a}^{\dagger}$. It is worth noting
at this juncture that the form of the $\CPn$ sigma model action 
(\ref{OperatorProjAction}) is
remarkably similar to the kinetic term of the static energy functional of 
a $(2+1)$-dimensional noncommutative scalar field (eq.(2.2) of ref\cite{GHS}).
Such noncommutative scalar field theories are known to exhibit a spectrum of 
localized field configurations (GMS solitons) \cite{GMS,GHS} with several interesting
properties. Not least among these is the rich structure of the $k$-soliton moduli 
space; a K\"ahler de-singularization of $(\R^{2})^{k}/S_{k}$, the symmetric 
product of the single soliton moduli space. It appears though, that this is not unique
to the scalar field theory \cite{Nekrasov1,Nekrasov2}. Indeed a similar resolution of 
the geometry of the $k$-soliton moduli space, as realized by the noncommutative 
algebra of projection operators, was demonstrated recently in the 
noncommutative $\CPn$ sigma model \cite{FurutaInami} by explicit computation of the
K\"ahler metric on the one- and two-soliton moduli space. In the light of such
remarkable evidence, it seems reasonable to ask if there may be further similarities
between GMS solitons and those of the $\CPn$ sigma model? With this in mind, it will
prove useful to proceed in close analogy with the analysis of GMS solitons.       
Returning to the $\CPn$ model and equating the energy of the configurations with the 
static action, it is 
easily seen that, as in the commutative case, the energy is bounded from below.
However, as will be demonstrated shortly, some degree of care must be exercised
when dealing with higher rank projectors\footnote{We thank Robert de
Mello Koch for drawing our attention to this point.}. 
Naively following \cite{GHS} it might seem like the energy may be written as
\begin{eqnarray}
 E_{\theta} = 2\pi {\rm Tr}_{\cal H}{\rm tr}
 {\Bigl(}2F_{\pm}(P)^{\dagger}F_{\pm}(P) \pm P{\Bigr)}\geq 2\pi 
 {\Bigl|}{\rm Tr}_{\cal H}{\rm tr}P{\Bigr|}\,,
 \label{BogBound}
\end{eqnarray}
\noindent
where 
\begin{eqnarray}
 F_{\pm}(P)=\left\{\begin{array}{l}
 (1-P)\widehat{a}P\\
 (1-P)\widehat{a}^{\dagger}P\end{array}\right.
\end{eqnarray}  
\noindent
The inequality would then saturate when the (anti)BPS equations 
$F_{\pm}(P) = 0$ are
satisfied and the topological charge of the BPS solitons takes 
on a particularly neat expression, being simply the combined matrix and Hilbert
space trace of the associated projector. However, this would be too naive! The
problem is that $\rm{Tr}_{\cal H}\rm{tr}(P)$ is generally infinite and hence
cannot represent the soliton charge. Crucial to the resolution of this issue is
the understanding that, in the noncommutative case, arguments in the trace may
{\it not} be permuted with impunity. With this in mind, returning to the 
static energy (\ref{OperatorProjAction}) 
(and focusing on the BPS case for the moment), 
it may be seen that\footnote{We would also like to thank O. Lechtenfeld for bringing to our attention ref.\cite{LP1,LP2} in which it was stressed that $E\ne{\rm Tr}_{\cal{H}}{\rm tr}P$ in the more general setting of a noncommutative $U(n)$-valued field in a modified $(2+1)$-dimensional sigma model.}
\begin{eqnarray}
 {\rm{Tr}}_{\cal H}{\rm{tr}}[P,\widehat{a}^{\dagger}]
 [\widehat{a},P] &=& {\rm{Tr}}_{\cal H}{\rm{tr}}
 {\Bigl(}P\widehat{a}^{\dagger}\widehat{a}P - P\widehat{a}^{\dagger}P
 \widehat{a} - \widehat{a}^{\dagger}P\widehat{a}P +
 \widehat{a}^{\dagger}P\widehat{a}{\Bigr)}\nonumber\\
 &=& {\rm{Tr}}_{\cal H}{\rm{tr}}{\Bigl(}2F_{+}(P)^{\dagger}F_{+}(P) + P -
 [\widehat{a},\widehat{a}^{\dagger}P] + 
 [P,\widehat{a}^{\dagger}P\widehat{a}]{\Bigr)}\,.
\end{eqnarray}  
\noindent
A straightforward computation shows that 
${\rm{Tr}}_{\cal H}{\rm{tr}}[P,\widehat{a}^{\dagger}P\widehat{a}]=0$ so that the 
last term may be dropped. Recalling the Weyl prescription mapping functions on a 
noncommutative space to an auxiliary Hilbert space, the second to last 
term may be thought of as an``integral of a derivative". 
As such, this may be evaluated with a noncommutative analogue of 
Stokes' theorem (see for instance \cite{Gross,Wadia})
\begin{eqnarray}
 {\rm{Tr}}_{\cal H}[\widehat{a},{\cal O}] =
 \lim_{M\rightarrow\infty}\sqrt{M+1}
 \langle M+1|{\cal O}|M\rangle
\end{eqnarray}
\noindent
where ${\cal O}$ is any appropriately well behaved operator on ${\cal H}$. 
For ${\cal O} = {\rm tr}\,\widehat{a}^{\dagger}P$, this term is generally
nonvanishing and cannot be neglected. With this in mind the energy functional
becomes 
\begin{eqnarray}
E_{\theta} &=& 2\pi{\rm{Tr}}_{\cal H}{\rm tr}{\Bigl(}
2F_{+}(P)^{\dagger}F_{+}(P)+ P
-[\widehat{a},\widehat{a}^{\dagger}P]{\Bigr)}\nonumber\\
&\geq& \underbrace{2\pi{\rm{Tr}}_{\cal H}{\rm tr}{\Bigl(} P - 
[\widehat{a},\widehat{a}^{\dagger}P]{\Bigl)}}_{2\pi Q_{+}}
\label{BogBound2}
\end{eqnarray}
\noindent
with $F_{+}(P)$ defined as above. Similarly it may be shown that
\begin{eqnarray}
E_{\theta} &=& 2\pi{\rm{Tr}}_{\cal H}{\rm tr}{\Bigl(}
2F_{-}(P)^{\dagger}F_{-}(P)- (P
-[\widehat{a},P\widehat{a}^{\dagger}]){\Bigr)}\nonumber\\
&\geq& \underbrace{2\pi{\Bigl|}{\rm{Tr}}_{\cal H}{\rm tr}{\Bigl(} P - 
[\widehat{a},P\widehat{a}^{\dagger}]{\Bigl)}{\Bigr|}}_{2\pi |Q_{-}|}
\label{BogBound3}
\end{eqnarray}
\noindent
Again, the energy bound is saturated for configurations for which the (anti)BPS
equations $F_{\pm}(P) = 0$ hold.
As shown in \cite{LeeLeeYang} such 
solutions are not hard to find; any Hermitian projector constructed from an
$(n+1)$-vector $W$ whose components are (anti)holomorphic polynomials will satisfy the
above (anti)BPS equations. These are the noncommutative extensions of the instanton
solutions of the conventional $\CPn$ sigma model. The static $1-$ and  $2$-soliton
solutions of the noncommutative $\CP$ model, for example, are given respectively by
\begin{eqnarray}
 W_{1} = \left(\begin{array}{c}
         a_{1}\\
	 \widehat{z} - b_{1}\end{array}\right)\,,\quad
 W_{2} = \left(\begin{array}{c}
         2a_{2}\widehat{z}+b_{2}\\
	 \widehat{z}^{2} + c_{2}\widehat{z} + e_{2}\end{array}\right)\,.	 	           
\end{eqnarray}
\noindent
The coefficients $a_{1},\ldots,e_{2}\in \C$ are chosen to 
coincide with the standard way of writing the
corresponding solitons of the commutative theory \cite{FurutaInami,Ward85}. These are
the complex moduli of the $\CPn$ instantons. The solutions are easily visualized in
the small $\theta$ limit by computing the energy density as an operator on the
auxiliary Hilbert space and mapping it back to a function on $\R^{2}_{\theta}$ by the
Weyl correspondence, {\it i.e.} 
$\widehat{\cal E}\mapsto{\cal E}_{\star} = {\cal W}^{-1}({\cal E})$. This is
exemplified by the simplest instanton solution \cite{LeeLeeYang}, 
$W_{1} = (1,\widehat{z})^{T}$ for which 
\begin{eqnarray}
  P = 
  \left( \begin{array}{cc}
  \frac{1}{1+\widehat{\bar{z}}\widehat{z}} &
  \frac{1}{1+\widehat{\bar{z}}\widehat{z}}\widehat{\bar{z}}\\
  \widehat{z}\frac{1}{1+\widehat{\bar{z}}\widehat{z}} & 
  \,\,\widehat{z}\frac{1}{1+\widehat{\bar{z}}\widehat{z}}\widehat{\bar{z}}
  \end{array} \right)\,.
  \label{CP1-1-solitonz}
\end{eqnarray}
\noindent
By way of illustration of the above points it is a useful exercise to compute the
topological charge of the above soliton\footnote{It is not too difficult to see 
that the instanton number must be independent of the
noncommutativity parameter so in computing $Q_{+}$, $\theta$ may be set to unity
without any loss of generality.}. The trace over ${\cal H}$ may be regulated
by 
the introduction of an infrared cutoff $M$ through the restriction 
to an $M$-dimensional subspace of ${\cal H}$ spanned by 
$\{|0\rangle,|1\rangle,\ldots,|M\rangle\}$ \cite{Gross}. As such
\begin{eqnarray}
 Q_{+} &=& \lim_{M\rightarrow\infty}\,\,{\rm Tr}_{{\cal
 H}_{M}}{\rm tr}P - {\rm Tr}_{{\cal H}_{M}}{\rm tr}[\widehat{a},
 \widehat{a}^{\dagger}P]\nonumber\\
 &=& \lim_{M\rightarrow\infty}\,\,\sum_{n=0}^{M}\langle n|\frac{1}{1+\widehat{N}}+
 \frac{1+\widehat{N}}{2+\widehat{N}}|n\rangle\nonumber\\
 &-& \sqrt{M+1}\langle
 M+1|\widehat{a}^{\dagger}\frac{1}{1+\widehat{N}}+
 \frac{\widehat{N}}{1+\widehat{N}}\widehat{a}^{\dagger}|M\rangle\nonumber\\
 &=& \lim_{M\rightarrow\infty}\,\,\sum_{n=0}^{M}{\Bigl(}\frac{1}{1+n}+
 \frac{1+n}{2+n}{\Bigr)} - {\Bigl(}1 + \frac{(M+1)^{2}}{M+2}{\Bigr)} = 1
 \label{One-inst-charge}
\end{eqnarray}
\noindent
In the small
$\theta$ limit, the energy density of the degree one instanton is written in terms of
the noncommuting coordinates on the plane as \cite{LeeLeeYang}
\begin{eqnarray}
 {\cal E}_{\star} = \frac{1}{1+\bar{z}\star z}
  \star\frac{1}{1+(\bar{z}\star z+\theta)} = \frac{1}{(1+\frac{1}
  {2}r^2)^2} + {\cal O}(\theta^2)\,.
\end{eqnarray}
\noindent
Note that the first corrections to the commutative instanton energy enter only at 
order $\theta^{2}$ so as long as $\theta$ is small the noncommutative instanton energy
density may be adequately approximated by the lowest order term in the $\theta$
perturbation series.
A similar computation for the degree two soliton $W_{2}$ (with $a_{2} = c_{2} = 0$)
gives the energy density
\begin{eqnarray}
 {\cal E}_{\star} =
  2\frac{|b_{2}|^{2}r^{2}}{(|b_{2}|^{2}+|e_{2}|^{2}+\frac{1}{4}r^{4}+
  e_{2}\bar{z}^{2}+
  \overline{e_{2}}z^{2})^{2}} + \cal{O}(\theta)\,.
\end{eqnarray}
\noindent
These solutions are plotted in fig.1 for various values of the complex moduli $b_{2}$
and $e_{2}$. It is interesting to note that the low energy scattering of the two
degree one instantons is not unlike that of the corresponding configuration of GMS
noncommutative scalar solitons \cite{GMS}. Indeed, this scattering property of was 
explicitly verified in the $\CP$ case in \cite{FurutaInami} where the metric on the 
two-soliton moduli space was directly computed.    
\FIGURE{\epsfig{file=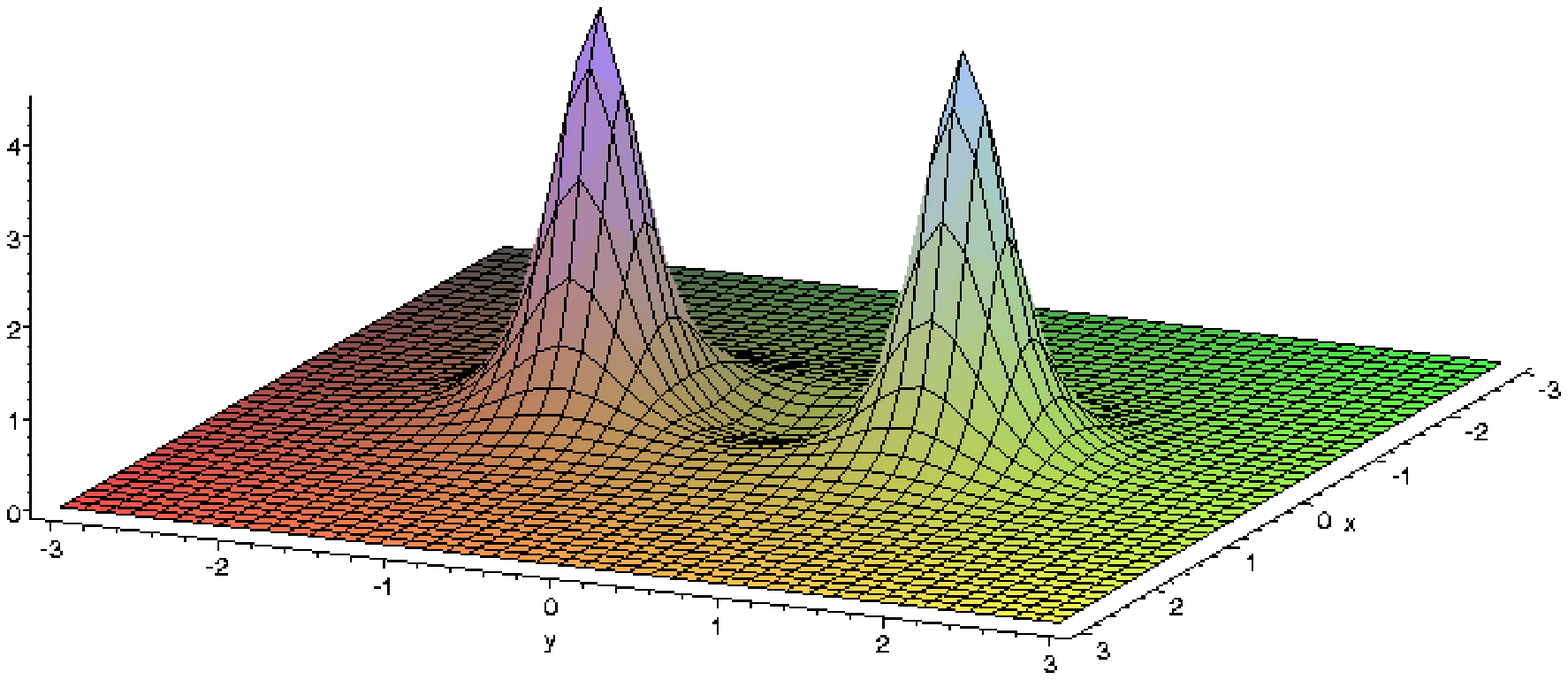,width=8cm,height=4.5cm}
         \epsfig{file=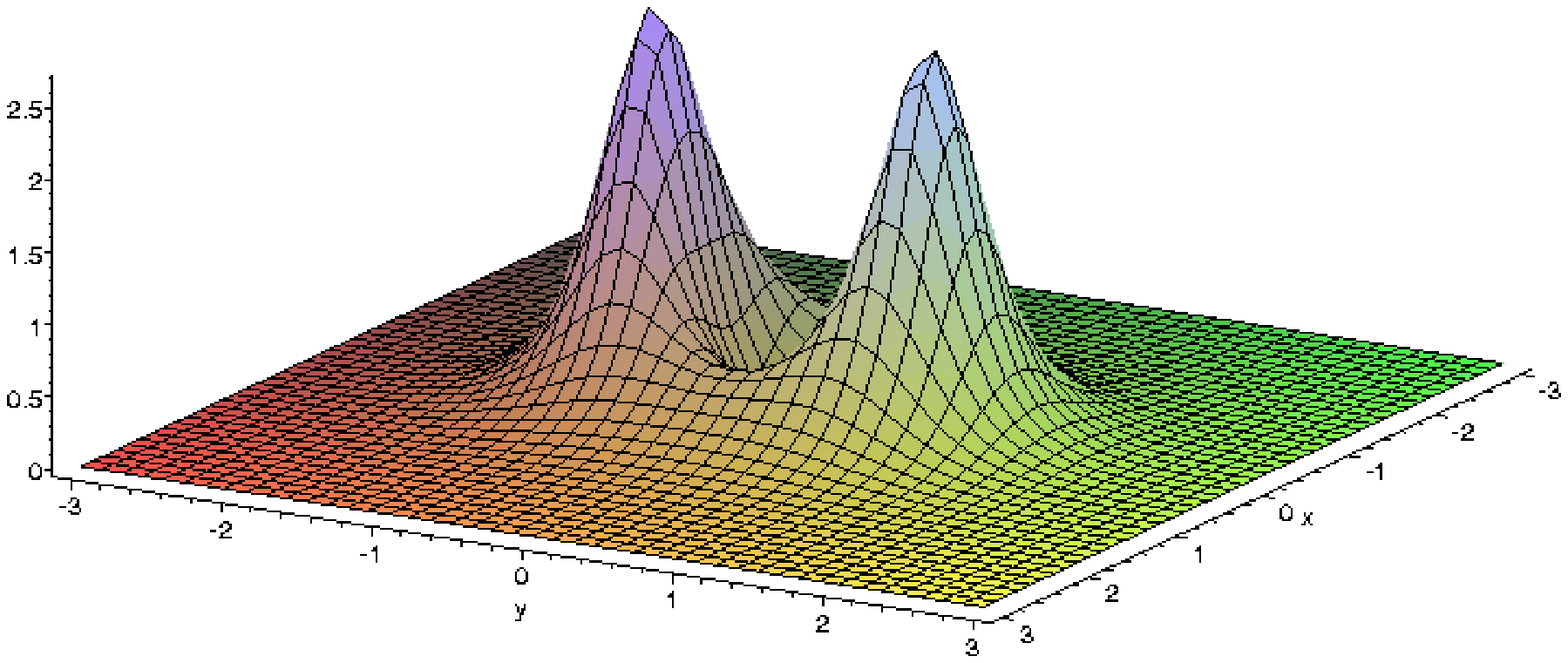,width=8cm,height=4.5cm}
	 \epsfig{file=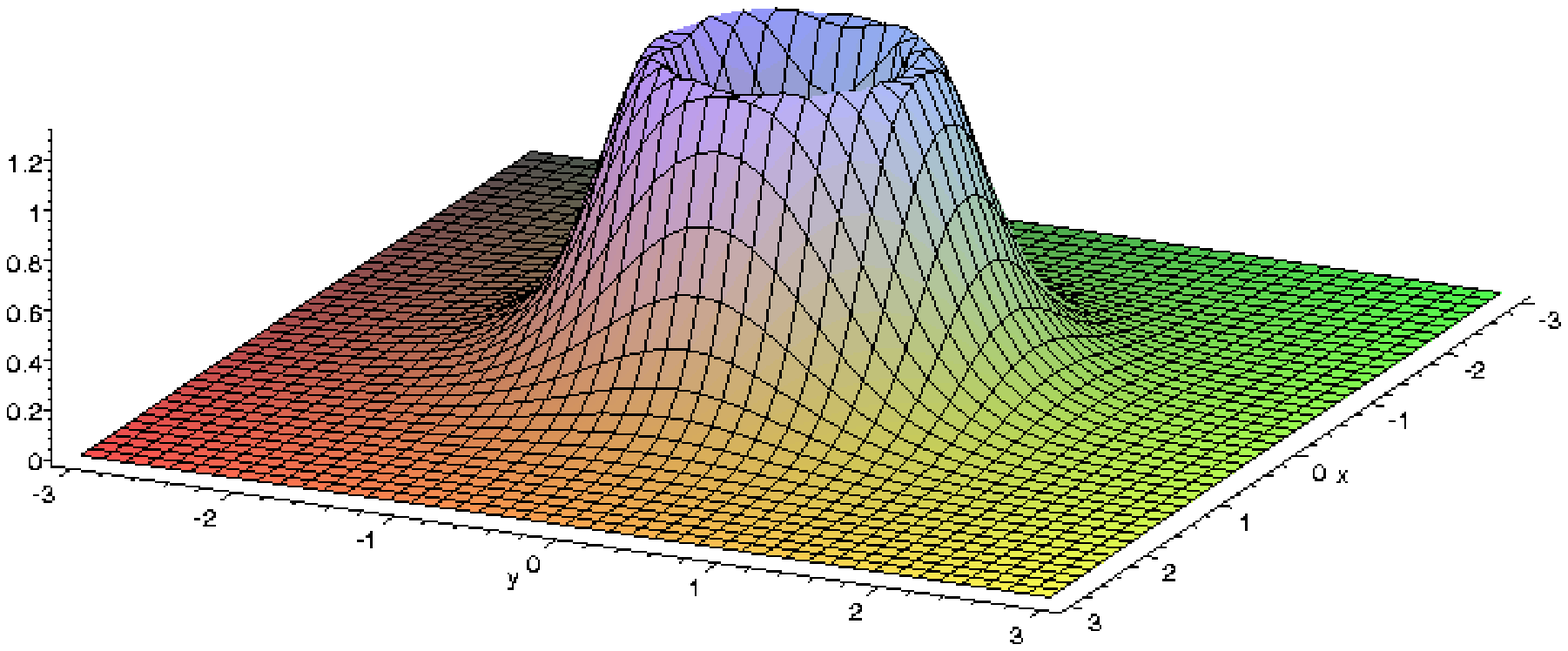,width=8cm,height=4.5cm}
 \caption{Static $2$-soliton solution of the $\CP$-model with complex moduli 
 $b_{2}=1$, and $e_{2}=1,0.5$ and $0$.}}

\section{Non-BPS States}
\label{Sec2}
\noindent
In addition to the simplest solutions of the $\CPn$ sigma model, the instanton
solutions described above; it is also known that these field theories (and their
generalizations to sigma models with Grassmannian target
$Gr(n,m) = SU(n+m)/S(U(n)\times U(m))$ possess a non-BPS sector consisting essentially
of bound states of instantons and anti-instantons 
\cite{DinZak1,DinZak2,Sasaki}. Such solutions solve the ($2nd$ order)
equations of motion without saturating any BPS bound on the energy functional and, not
protected by supersymmetry, are in general unstable. 
These classical solutions have also resurfaced recently \cite{Ioannidou} when it 
was shown that not only do they solve the $\CPn$ sigma model equations but that they
also solve a Dirac-Born-Infeld (DBI) type action pointing to a bulk interpretation of
these solitons as D-brane states although the precise states that they correspond to is not
yet clear.\\

\subsection{Constructing non-BPS states - The commutative case}
\noindent
In this section we aim to continue the
analysis carried out in \cite{Sasaki} and ask if such non-BPS states persist when the
base space of the sigma model is made noncommutative. To this end, we briefly review
the elegant construction employed in \cite{Sasaki}, modifying it to explicitly treat 
the $\CPn$ sigma model. The idea behind said approach is rather elementary, demanding
only a little linear algebra. Given a holomorphic 
$(n+1)$-vector ${\bf f}$ which characterizes the $\CPn$ instanton, a set of $n+1$ vectors 
$\{{\bf f}_1,{\bf f}_2,\ldots,{\bf f}_{n+1}\}$ is constructed from ${\bf f}$ (as described
below) such that $span\{{\bf f}_1,{\bf f}_2,\ldots,{\bf f}_{n+1}\} = \C^{n+1}$. 
This set may then be
orthonormalized by the conventional Gram-Schmidt procedure and, remarkably, any vector
in the resulting orthonormal set is a solution of the $\CPn$ equations of motion.
Indeed, this may be seen quite easily as follows;
in terms of the complex coordinates $(z,{\bar z})$ on $\R^{2}$ and 
the Hermitian projector $P$,
the $\CPn$ sigma model equations of motion may be written as
\begin{eqnarray}
  [\partial_{z} \partial_{\bar{z}}P,P] = 0\,.
  \label{ELProj}
\end{eqnarray}
\noindent
Let ${\bf f}$ be some holomorphic $(n+1)$ component vector (any instanton solution will do)
and define ${\bf f}_{1}:={\bf f}, {\bf f}_{2}:=\partial_{z}{\bf f},\ldots, 
{\bf f}_{n+1}:=\partial^{n}_{z}{\bf f}$. Assuming linear independence of the 
${\bf f}_{i}$ for $1\leq i\leq n+1$ means that they
span $\C^{n+1}$. This set may be orthonormalized by the usual Gram-Schmidt procedure to
give an orthonormal basis for $\C^{n+1}$ as follows: Define ${\bf e}'_{1}:={\bf f}_{1}$,
${\bf e}_{1}:={\bf e}'_{1}/({\bf e}'^{\dagger}_{1}\cdot {\bf e}'_{1})^{1/2}$ and  
\begin{eqnarray}
 {\bf e}'_{i}&:=& {\bf f}_{i} - \sum_{j=1}^{i-1}\, {\bf e}_{j}({\bf e}_{j}^{\dagger}
 \cdot {\bf f}_{i})\nonumber\\
 {\bf e}_{i}&:=&\frac{{\bf e}'_{i}}{({\bf e}'^{\dagger}_{i}\cdot {\bf e}'_{i})^{1/2}}\,,
 \label{Gram-Schmidt}
\end{eqnarray}
\noindent
for $2\leq i \leq n+1$. It then follows quite straightforwardly that
\begin{eqnarray}
  P_{i} := {\bf e}_{i} {\bf e}_{i}^{\dagger}\quad 1 \leq i \leq n+1
  \label{non-BPS-proj}
\end{eqnarray}
\noindent
is a Hermitian projector. To show that the $\{{\bf e}_{i}\}$ form a set of solutions of the
$\CPn$ model, it suffices to show that the $P_{i}$ solve the sigma-model equations of
motion (\ref{ELProj}) for any $1\leq i\leq n+1$. To this end, it will prove useful to define
the auxiliary matrix variable
\begin{eqnarray}
  Q_{i} := \sum_{j=1}^{i-1}\,{\bf e}_{j} {\bf e}_{j}^{\dagger}\,.
\end{eqnarray}
\noindent
Clearly $Q$ is also a Hermitian projection operator orthogonal to $P$ 
since (for fixed $i$) $P_{i}Q_{i} = Q_{i}P_{i} = 0$. A few lines of algebra
together with the identities
\begin{eqnarray}
  \partial_{\bar{z}}{\bf e}_{i} &=& \sum_{k=1}^{i}\,{\bf e}_{k}({\bf e}_{k}^{\dagger}
  \partial_{\bar{z}}{\bf e}_{i})\nonumber\\
  \partial_{z}{\bf e}_{i} &=& \sum_{k=1}^{i+1}\,{\bf e}_{k}({\bf e}_{k}^{\dagger}
  \partial_{z}{\bf e}_{i})\,,
  \label{deriv-identities}
\end{eqnarray}
\noindent
establishes that\footnote{For a detailed derivation of these properties of the
projection operators we refer the interested reader to {\cite{Sasaki}} and
relevant references therein} $\partial_{\bar{z}}Q_{i}Q_{i} = 
P_{i}\partial_{\bar{z}}Q_{i} = \partial_{\bar{z}}(P_{i}+Q_{i})(P_{i}+Q_{i}) = 
\partial_{\bar{z}}P_{i}Q_{i} + P_{i}\partial_{\bar{z}}Q_{i} = 0$. The last equality of
course follows from differentiation of the orthogonality relation satisfied by the
$P_{i}$'s and $Q_{i}$'s. In much the same way it is also easy to verify that
$P_{i}\partial_{z} Q_{i} = \partial_{z} Q_{i}$. Combining these gives
\begin{eqnarray}
  \partial_{\bar{z}}P_{i}P_{i} + \partial_{\bar{z}}Q_{i} = 0\,,
\end{eqnarray}
\noindent
and by Hermitian conjugation
\begin{eqnarray}
  P_{i} \partial_{z} P_{i} + \partial_{z} Q_{i} = 0\,.
\end{eqnarray} 
\noindent
Taking the holomorphic derivative of the former and subtracting the
antiholomorphic derivative of the latter gives the desired commutator and
completes the proof. In the (commutative) classical $\CPn$ sigma model, this
procedure can be shown {\cite{DinZak1,DinZak2}} to generate 
{\it the most general} finite action solutions of the sigma model equations 
of motion. These solutions are interpreted variously as instantons, 
anti-instantons or unstable noninteracting mixtures thereof. 
In recent work {\cite{Ioannidou}} it was
also shown that these non-BPS solitons of the $\CPn$ model are not only finite
action solutions of the sigma model but are also finite action solutions of a
Dirac-Born-Infeld (DBI) model with a $\CPn$ target space.\\

\subsection{Constructing non-BPS states - The noncommutative case}
\noindent
Our focus is however on the noncommutative theory and, as such, one question of
interest is whether or not the non-BPS construction above extends to the
noncommutative $\CPn$ model. Passing to the noncommutative variables 
$\widehat{z}$ and $\widehat{\bar{z}}$, 
results in the equations of motion 
\begin{eqnarray}
 {\Bigl[}[\widehat{\bar{z}},[\widehat{z},P]],P{\Bigr]}=0\,,
 \label{NCELEQ}
\end{eqnarray}
while the BPS and anti-BPS equations are, respectively
\begin{eqnarray}
 (1-P)\widehat{z}P &=& 0\nonumber\\
 (1-P)\widehat{\bar{z}}P &=& 0\,.
 \label{NCBPSEQ}
\end{eqnarray}  
Any solution of the (anti-)BPS equations is also a solution of the Euler-Lagrange
equations of motion; a fact that is obvious when the latter is written as
$[\widehat{\bar{z}},(1-P)\widehat{z}P] + [\widehat{z},P\widehat{\bar{z}}(1-P)] = 0$ or equivalently
 $[\widehat{z},(1-P)\widehat{\bar{z}}P] +
[\widehat{\bar{z}},P\widehat{z}(1-P)] = 0$. 
The reverse is, of course certainly not true in general and
solutions of (\ref{NCELEQ}) (if they exist) that do not solve 
(\ref{NCBPSEQ}) are precisely the non-BPS states. We now attempt to find such
solutions by adapting the orthonormaization construction of {\cite{Sasaki}}.
\noindent
Let $W$ be a holomorphic $(n+1)$-vector and define 
\begin{eqnarray}
 {\bf f}_{1}&:=&W\nonumber\\ 
 {\bf f}_{2}&:=&-\frac{1}{\theta}[\widehat{\bar{z}},W]\nonumber\\
 &\vdots&\nonumber\\
 {\bf f}_{k}&:=&(-1)^{k-1}\frac{1}{\theta^{k-1}}\underbrace{
 [\widehat{\bar{z}},\dots,[\widehat{\bar{z}},W]\dots]}_{(k-1)\,\,\rm{commutators}}\nonumber\\
 &\vdots&
\end{eqnarray}
\noindent
The set $\{{\bf f}_{1},{\bf f}_{2},\ldots,{\bf f}_{n+1}\}$ is
orthonormalized as follows: Choose ${\bf e}_{1} = W(1/\sqrt{W^{\dagger}W})$ and
write
\begin{eqnarray}
 {\bf e'}_{2} &=& {\bf f}_{2} - {\bf e}_{1}({\bf e}_{1}^{\dagger}{\bf f}_{2}) 
 = -\frac{1}{\theta}{\Bigl\{}[\widehat{\bar{z}},W] - W\frac{1}{W^{\dagger}W}W^{\dagger}
 [\widehat{\bar{z}},W]{\Bigr\}}\nonumber\\
 &=& -\frac{1}{\theta}(1-P_{1})\widehat{\bar{z}}W\,,
\end{eqnarray}
\noindent
where $P_{1} = W(W^{\dagger}W)^{-1}W^{\dagger}$ is a Hermitian projection operator
and in the last line, use was made of the fact that $W$ is an eigenvector of $P_{1}$
with unit eigenvalue. Computing the norm of ${\bf e'}_{2}$ as
$(1/\theta)W^{\dagger}\widehat{z}(1-P_{1})\widehat{\bar{z}}W$ allows us to write
\begin{eqnarray}
  {\bf e}_{2} &=& -(1-P_{1})\widehat{\bar{z}}W\frac{1}
  {\sqrt{W^{\dagger}\widehat{z}(1-P_{1})\widehat{\bar{z}}W}}\,.
  \label{OrthoBasis}
\end{eqnarray}  
\noindent
with the associated Hermitian projection operator
\begin{eqnarray}
 P_{2}=
 (1-P_{1})\widehat{\bar{z}}W\frac{1}
 {W^{\dagger}\widehat{z}(1-P_{1})\widehat{\bar{z}}W}
 W^{\dagger}\widehat{z}(1-P_{1})\,.
 \label{Projectionop2}
\end{eqnarray}
\noindent
As in the commutative case, defining $P_{j}:= {\bf e}_{j}{\bf e}_{j}^{\dagger}$ as the
Hermitian projector associated to the $j$'th (orthonormal) basis vector we can by iteration
construct
\begin{eqnarray}
 {\bf e'}_{k} &:=& (-1)^{k-1}\frac{1}{\theta^{k-1}}{\Bigl(}1-\sum_{j=1}^{k-1}\,P_{j}{\Bigr)}
 {\Bigl[}\widehat{\bar{z}},\dots,[\widehat{\bar{z}},W]\dots{\Bigr]}\nonumber\\
 &=& (-1)^{k-1}\frac{1}{\theta^{k-1}}{\Bigl(}1-\sum_{j=1}^{k-1}\,P_{j}{\Bigr)}
 \widehat{\bar{z}}^{k-1}W\,,
 \label{ek}
\end{eqnarray}
\noindent
where the last equality follows iteratively from the fact that $W\in\,{\rm ker}(1-P_{1})$ and
${\bf e}_{k}$ is constructed from ${\bf e'}_{k}$ by the usual normalization.
We leave it as a trivial exercise to the reader to verify that the set 
$\{{\bf e}_{1},\dots,{\bf e}_{n+1}\}$ is indeed orthonormal. That ${\bf e}_{k}$ as
constructed above solves the sigma model equations of motion follows in close analogy
to the commutative case. For concreteness though, we show this explicitly for the case $k=2$.
Observe that the projection operators $P_{1}$ and $P_{2}$ satisfy the relation
\begin{eqnarray}
 [\widehat{z},P_{2}]P_{2} + [\widehat{z},P_{1}]P_{2} = 0\,.
 \label{Prop1}
\end{eqnarray}
\noindent
Moreover, the commutative derivative relation $\partial_{z}P_{1} = {\bf e}_{2}{\bf
e}_{2}^{\dagger}(\partial_{z}{\bf e}_{1}){\bf e}_{1}^{\dagger}$ translates in
noncommutative coordinates to $[\widehat{\bar{z}},P_{1}] = 
P_{2}\widehat{\bar{z}}P_{1}$ so that $P_{1}$ and $P_{2}$ further satisfy
\begin{eqnarray}
 P_{2}[\widehat{\bar{z}},P_{1}] &=& [\widehat{\bar{z}},P_{1}]\,.
 \label{Prop2}
\end{eqnarray}
\noindent
Substituting this into eq.(\ref{Prop1}) and applying the commutator 
$[\widehat{\bar{z}},\cdot]$ to the resulting equation gives
\begin{eqnarray}
 {\Bigl[}\widehat{\bar{z}},[\widehat{z},P_{2}]P_{2}{\Bigr]} &+& 
 {\Bigl[}\widehat{\bar{z}},[\widehat{z},P_{1}]{\Bigr]}\nonumber\\
 &=& 
 [\widehat{z},P_{2}][\widehat{\bar{z}},P_{2}] + {\Bigl[}\widehat{\bar{z}}
 ,[\widehat{z},P_{2}]{\Bigr]}P_{2} + 
 {\Bigl[}\widehat{\bar{z}},[\widehat{z},P_{1}]{\Bigr]}= 0\,,
 \label{Prop3}
\end{eqnarray} 
where in the last step, we have made use of the Jacobi identity $[A,[B,C]]$ + 
$cyclic$ $permutations$ $=0$ and the Heisenberg algebra satisfied by the 
noncommuting coordinates. In the latter form it is clear that the first and third
terms in (\ref{Prop3}) are self-adjoint and so the subtraction from (\ref{Prop3}) 
of its Hermitian conjugate shows that $P_{2}$ satisfies (\ref{NCELEQ}) and verifies
our claim that ${\bf e}_{2}$ is, in fact an exact solution of the noncommutative 
$\CP$ sigma model. That this is true, in itself should not be surprising given our
construction. A further computation shows that  
\begin{eqnarray}
 (1-P_{2})\widehat{\bar{z}}P_{2} = [\widehat{\bar{z}},(P_{1}+P_{2})]\,.
 \label{Non-antiBPS}
\end{eqnarray}
\noindent
If the commutator on the right hand side of eq.(\ref{Non-antiBPS}) vanishes can we conclude
that ${\bf e}_{2}$ is a {\it non-BPS} soliton\footnote{This is slightly abusive
terminology since (\ref{Non-antiBPS}) would ensure only that ${\bf e}_{2}$ is not
an anti-BPS soliton. We shall take ``non-BPS" to mean both equations in 
(\ref{NCBPSEQ}) are nonvanishing.}. \\

\subsection{Examples}
\noindent
This noncommutative modification of the Sasaki-Din-Zakrewski (SDZ) construction 
is perhaps best illustrated by some examples.  
\begin{itemize}
\item $\underline{\CP}$: To begin with, let us consider the case $n=1$.
It is a well known fact \cite{DinZak1,DinZak2,ZakBook} that for the commutative $\CP$ sigma
model the SDZ construction maps instantons directly to their corresponding anti-instanton
solutions.
Since the construction yields a complete set of finite action solutions to the sigma model
equations of motion it follows then that the commutative $\CP$ sigma model {\it does not
possess a non-BPS spectrum}. One might naturally ask if the same is true for the
noncommutative $\CP$ sigma model. It was already shown in \cite{LeeLeeYang} that 
the simplest BPS solution of the noncommutative $\CP$ model is the $Q=1$ instanton 
with $W=(1,z)^{T}$ and associated projector (\ref{CP1-1-solitonz}).
Substituting this into the expression for ${\bf e}_{2}$ in (\ref{OrthoBasis}),
simplifying the resulting $2$-vector and relabelling the solitonic configuration
by $\widetilde{W_{2}}$ we get
\begin{eqnarray}
  \widetilde{W_{2}} =
     \left( \begin{array}{c}
      -\frac{1}{1+\widehat{\bar{z}}\widehat{z}}\widehat{\bar{z}}\\
      \frac{1}{1+\widehat{\bar{z}}\widehat{z}} \end{array}\right)
      \sqrt{1+\widehat{\bar{z}}\widehat{z}}= \left( \begin{array}{c}
      -\widehat{\bar{z}}\\
      1 \end{array}\right)\frac{1}{\sqrt{1+\widehat{\bar{z}}\widehat{z}}}\,.  
\end{eqnarray}
\noindent
This is precisely the normalized anti-holomorphic vector corresponding to
the anti-instanton solution expected of the SDZ
construction for $\CP$. This is easily verified by noting that $P_{1}+P_{2} = 1\!\!1$ 
so that the commutator on the right hand side of eq.(\ref{Non-antiBPS}) vanishes. However,
concluding from this that, as in the commutative case, the noncommutative $\CP$ sigma model
does not possess a non-BPS sector would be at best premature (and certainly in this case 
erroneous). In a remarkable recent work \cite{FurutaInami2} a large class of non-BPS 
configurations were
constructed from meta-stable bound states of solitons and anti-solitons of the
{\it GMS} noncommutative scalar field theory \cite{GMS},\cite{GHS}. The
construction of \cite{FurutaInami2} hinges on the fact that in a basis that
diagonalizes the $(2\times 2)$ Hermitian projector $P$ associated to a solution
of the $\CP$ sigma model equations, the diagonal entries 
$\phi_{1}(\widehat{z},\widehat{\bar{z}})$ and $\phi_{2}(\widehat{z},\widehat{\bar{z}})$ 
will also solve (\ref{NCELEQ}).
In particular if $\phi_{1}$($\phi_{2}$) are taken to be GMS (anti)solitons
satisfying $(1-\phi_{1})\widehat{z}\phi_{1}=0$ and 
$(1-\phi_{2})\widehat{\bar{z}}\phi_{2}=0$ respectively, then $P$ does not solve
either of the equations in (\ref{NCBPSEQ}) and the corresponding field
configuration $W$ is a non-BPS soliton of the $\CP$ sigma model. From this example it is
alarmingly clear that the SDZ construction does not saturate the set of solutions of the
noncommutative $\CP$ sigma model.

\item
$\underline{\C\Pk^{2}}$: Having shown that the modified SDZ construction is insensitive 
to the non-BPS spectrum of the noncommutative $\CP$ sigma model we now consider the $n=2$
case. Analysis of these solutions will prove useful in facilitating comparison with the work
of \cite{FurutaInami2}. The simplest instanton solution of the $\C\Pk^{2}$ sigma model is
$W=(1,\widehat{z},\widehat{z}^2)^{T}$. The corresponding Hermitian projector is computed to be 
\begin{eqnarray}
 P_{1}:= W\frac{1}{W^{\dagger}W}W^{\dagger} = \left( \begin{array}{ccc}
  \frac{1}{A} & \frac{1}{A}\widehat{\bar{z}} & \frac{1}{A}\widehat{\bar{z}}^{2}\\
  \widehat{z}\frac{1}{A} & \widehat{z}\frac{1}{A}\widehat{\bar{z}} & \widehat{z}
  \frac{1}{A}\widehat{\bar{z}}^{2}\\
  \widehat{z}^2\frac{1}{A} & \widehat{z}^2\frac{1}{A}\widehat{\bar{z}} 
  & \widehat{z}^2\frac{1}{A}\widehat{\bar{z}}^{2}
  \end{array} \right)\,,
\end{eqnarray} 
\noindent
where $A(\widehat{\bar{z}}\widehat{z}) = 1 + \widehat{\bar{z}}\widehat{z} 
- \theta\widehat{\bar{z}}\widehat{z} + (\widehat{\bar{z}}\widehat{z})^{2}$ 
is the square modulus of
$W$. Using the relations $\widehat{\bar{z}}f(\widehat{\bar{z}}\widehat{z}) 
= f(\widehat{\bar{z}}\widehat{z} - \theta)\widehat{\bar{z}}$ and
$\widehat{z}f(\widehat{\bar{z}}\widehat{z}) = 
f(\widehat{\bar{z}}\widehat{z}+\theta)\widehat{z}$ we find
\begin{eqnarray}
  \widetilde{W_{2}}&:=&{\bf e}_{2} = -(1-P_{1})\widehat{\bar{z}}W
  \frac{1}{\sqrt{W^{\dagger}\widehat{z}(1-P_{1})\widehat{\bar{z}}W}}\nonumber\\
 &=& \left( \begin{array}{c}
      -\widehat{\bar{z}}(1+2\widehat{\bar{z}}\widehat{z})\\
      1 - \theta\widehat{\bar{z}}\widehat{z} - (\widehat{\bar{z}}\widehat{z})^{2}\\
      \widehat{z}(\widehat{z}\widehat{\bar{z}} + 2) 
      \end{array}\right)
      \frac{1}{\sqrt{B(\widehat{\bar{z}}\widehat{z})}}\,,
\end{eqnarray}
\noindent
with $B(\widehat{\bar{z}}\widehat{z}) = 1+\theta 
+ (5+6\theta+\theta^{2})\widehat{\bar{z}}\widehat{z} +
(6+6\theta+\theta^{2})(\widehat{\bar{z}}\widehat{z})^{2} 
+ (5+2\theta)(\widehat{\bar{z}}\widehat{z})^{3} 
+ (\widehat{\bar{z}}\widehat{z})^{4}$. It is
straightforward (but tedious) to compute $P_{2} = 
\widetilde{W_{2}}^{\dagger}\widetilde{W_{2}}$ and check that the commutator 
$[\widehat{\bar{z}},(P_{1}+P_{2})]$ is nonvanishing and so conclude that 
$\widetilde{W_{2}}$ is a genuine non-BPS soliton of the $\C\Pk^{2}$ sigma model. As a check
we find that in the $\theta \rightarrow 0$ limit $\widetilde{W_{2}}$ becomes
\begin{eqnarray}
 \left( \begin{array}{c}
      -\bar{z}(1 + r^{2})\\
      1 - \frac{1}{4}r^{4}\\
      z(\frac{1}{2}r^{2} + 2) 
      \end{array}\right)\frac{1}{\sqrt{1+\frac{5}{2}r^{2}+\frac{3}{2}r^{4}
      +\frac{5}{8}r^{6}+\frac{1}{16}r^{8}}}\,,
\end{eqnarray}
\noindent
in complete agreement with \cite{DinZak1}. 
\item $\underline{\C\Pk^{2}}$:
As a final illustration of the construction
technique we start with the $\C\Pk^{2}$ instanton 
$W = (\widehat{z}^2+1,\widehat{z}^2-1,2\widehat{z})^{T}$. With 
$A(\widehat{\bar{z}}\widehat{z}) = 2+ (4-2\theta)\widehat{\bar{z}}\widehat{z} 
+ 2(\widehat{\bar{z}}\widehat{z})^{2}$, the
corresponding projection operator is written
\begin{eqnarray}
 P_{1}= \left( \begin{array}{ccc}
  (\widehat{z}^{2}+1)\frac{1}{A}(\widehat{\bar{z}}^{2}+1) 
  & \,\,(\widehat{z}^{2}+1)\frac{1}{A}(\widehat{\bar{z}}^{2}-1) & 
  \,\,2(\widehat{z}^{2}+1)\frac{1}{A}\widehat{\bar{z}}\\
  (\widehat{z}^{2}-1)\frac{1}{A}(\widehat{\bar{z}}^{2}+1) 
  & \,\,(\widehat{z}^{2}-1)\frac{1}{A}(\widehat{\bar{z}}^{2}-1) & 
  \,\,2(\widehat{z}^{2}-1)\frac{1}{A}\widehat{\bar{z}}\\
  2\widehat{z}\frac{1}{A}(\widehat{\bar{z}}^{2}+1)
  & \,\,2\widehat{z}\frac{1}{A}(\widehat{\bar{z}}^{2}-1)& 
  \,\,4z\frac{1}{A}\widehat{\bar{z}}
  \end{array} \right)\,.
\end{eqnarray}
\noindent
If we define $B(\widehat{\bar{z}}\widehat{z}) = 1 + 2\theta 
+ (4+6\theta +2\theta^{2})\widehat{\bar{z}}\widehat{z} + (6 + 6\theta +
\theta^{2})(\widehat{\bar{z}}\widehat{z})^{2} 
+ (4 + 2\theta)(\bar{z}z)^{3} + (\widehat{\bar{z}}\widehat{z})^{4}$ then the
(normalized) non-BPS
soliton constructed from $W$ may be written
\begin{eqnarray}
 \widetilde{W_{2}} = \left[\left( \begin{array}{c}
      \widehat{z}-\widehat{\bar{z}}\\
      \widehat{z}+\widehat{\bar{z}}\\
      1-\widehat{\bar{z}}\widehat{z}
      \end{array}\right)(1+\widehat{\bar{z}}\widehat{z}) + 
      \theta \left( \begin{array}{c}
      \widehat{z}\\
      \widehat{z}\\
      -\widehat{z}\widehat{\bar{z}}
      \end{array}\right)\right]\frac{1}{\sqrt{B(\widehat{\bar{z}}\widehat{z})}}\,.
 \label{CP2-nonBPS2}     
\end{eqnarray}
\noindent
In this form the commutative limit is very conveniently investigated. Sending $\theta$ to
zero and noticing that $B(\widehat{\bar{z}}\widehat{z}) \rightarrow (1+\bar{z}z)^{4}$ 
reduces 
$\widetilde{W_{2}}$ to 
\begin{eqnarray}
     \left( \begin{array}{c}
      z-\bar{z}\\
      z+\bar{z}\\
      1-\frac{1}{2}r^{2} \end{array}\right)\frac{1}{1+\frac{1}{2}r^{2}}\,.      
\end{eqnarray}
Again, this corresponds exactly to what is expected in the commutative limit. As can be seen
from the soliton energy density (see Fig.2) the non-BPS state is formed from a bound state
of two degree-$1$ $\C\Pk^{2}$ instantons and two anti-instantons all coincident at the 
origin.
\end{itemize}

\FIGURE{\epsfig{file=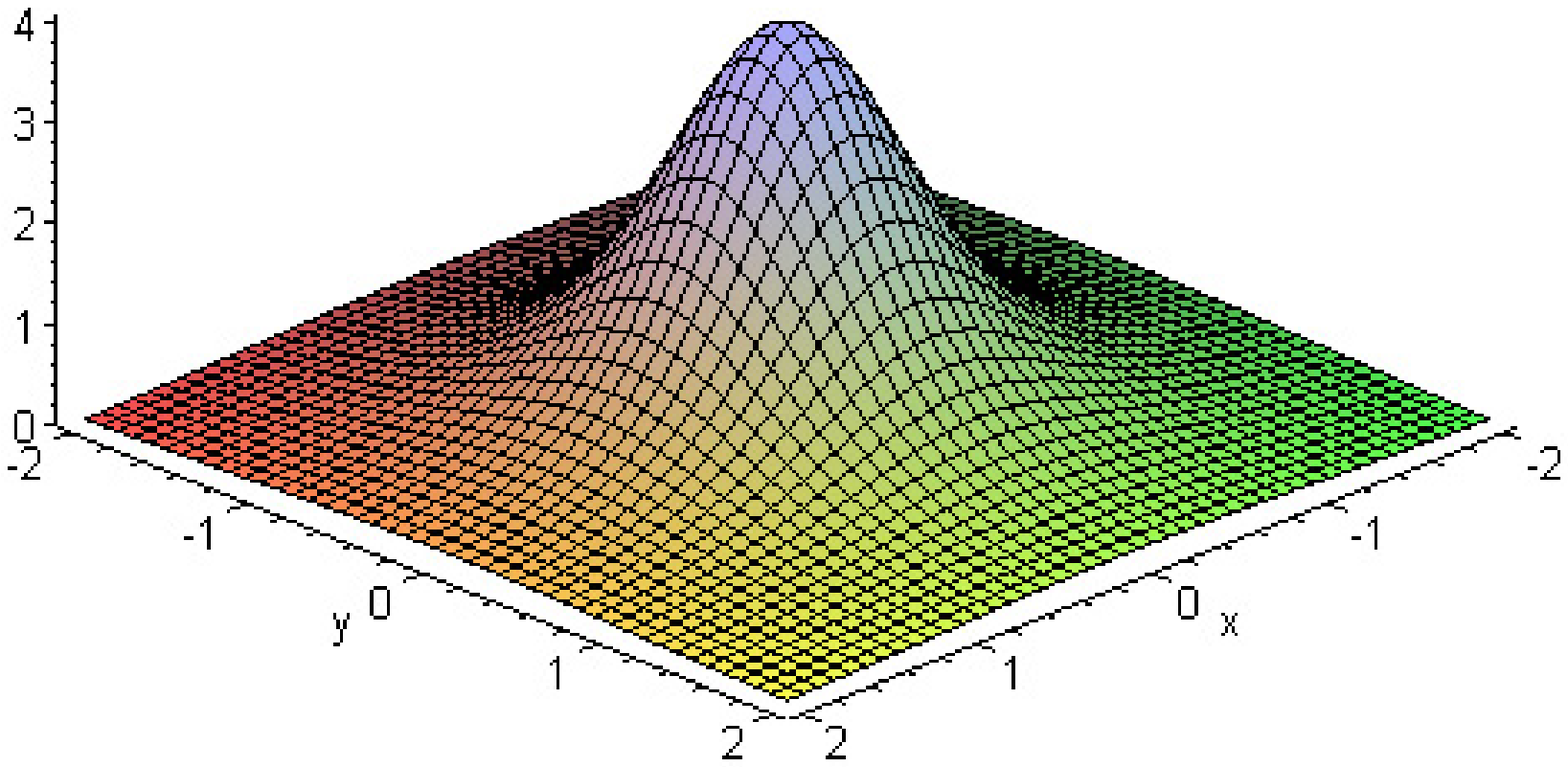,width=4cm,height=3cm}
 \caption{A $\C\Pk^{2}$ bound state consisting of two instantons and two anti-instantons all coincident at the origin with total energy density four times that of a single instanton.}}   

\section{$\CPn$ solitons and GMS solitons}
\label{Sec3}
\noindent
The existence of a non-BPS spectrum of the noncommutative $\CP$ model is
intruging \cite{FurutaInami2}; even
more so since the contruction of non-BPS solitonic confugurations is so
intimately connected to the noncommutative scalar solitons of \cite{GMS}. It seems
only natural then to try and probe this connection further in the hope of a deeper
understanding of the space of solutions to the noncommutative $\CPn$ sigma model.
Returning to the BPS solitons of Sec.(2.3), it may be immediately seen that an
alternative construction of degree $k$ solitons would be to take the Hermitian
projector $P$ in the diagonal representation (as in \cite{FurutaInami2})     
\begin{eqnarray}
 P = \left( \begin{array}{cccc}
  \widehat{\phi}_{1} & 0 & \ldots & 0\\
  0 & \widehat{\phi}_{2} & 0\,\,\ldots & 0\\
  \vdots & \vdots & \vdots & \vdots\\
  0 & \ldots & 0 & \widehat{\phi}_{n+1}
  \end{array} \right)\,,
  \label{DiagP}
\end{eqnarray}
\noindent
where each of the $\widehat{\phi}$'s satisfy\footnote{For
concreteness, we shall focus only on the BPS solutions with the understanding that
similar relations hold for the anti-BPS case.}
\begin{eqnarray}
 (1-\widehat{\phi}_{i})\widehat{a}\widehat{\phi}_{i} = 0
 \label{GMSeqn}
\end{eqnarray}
and $\sum_{i}{\rm Rank}(\widehat{\phi}_{i}) = k$. This necessarily implies that
$(1-P)\widehat{a}P = 0$ -- precisely the condition that $P$ corresponds to a BPS
$\CPn$ configuration. In general, solutions to (\ref{GMSeqn}) are parameterized by
$k_{i}$ complex coherent state vectors $|z_{i}^{a}\rangle := {\rm
exp}(z_{i}^{a}\widehat{a}^{\dagger})|0\rangle$ with 
\begin{eqnarray}
 \widehat{\phi}_{i} = \sum_{a,b=1}^{k_{i}}\,|z_{i}^{a}\rangle
 \frac{1}{\langle z_{i}^{a}|z_{i}^{b}\rangle}\langle z_{i}^{b}|
\end{eqnarray}
\noindent
These are, of course, the noncommutative scalar solitons of the GMS
model\cite{GMS,GHS}. Apparently then, in addition to the standard construction of
$k$-soliton solutions of the noncommutative $\CPn$ sigma model
\cite{LeeLeeYang,FurutaInami,FurutaInami2}, such solutions may also be constructed
from stacking GMS solitons of appropriate charge on the plane. The
emergence of a non-BPS sector of the $\CP$ model for finite $\theta$ via such a
construction lends much weight in favour of this claim. It will be shown in this
section however that, at least for (anti-)BPS solutions, such an `alternative' 
construction should have been
expected since (\ref{DiagP}) is just the diagonal represention of the usual
Hermitian projector associated to BPS solutions of the $\CPn$ sigma model. This is
most easily illustrated for the case of the static $1$-soliton solution of the
noncommutative $\CP$ model for which $P$ is given by
\begin{eqnarray}
  P=\left( \begin{array}{cc}
  \frac{1}{1+\widehat{N}} &
  \frac{1}{1+\widehat{N}}\widehat{a}^{\dagger}\\
  \widehat{a}\frac{1}{1+\widehat{N}} &
  \,\,\frac{1+\widehat{N}}{2+\widehat{N}}
  \end{array} \right)
  \label{CP1-1-solitona}
\end{eqnarray}
\noindent
after setting $\theta = 1$. Denoting $|\One\rangle = (1,0)^{T}$ and $|\Two\rangle =
(0,1)^{T}$,
an eigenvector of $P$ with eigenvalue $\lambda$ may be expanded as
\begin{eqnarray}
|\Psi\rangle = |\psi_{1}\rangle\otimes |\One\rangle + 
|\psi_{2}\rangle\otimes |\Two\rangle
\end{eqnarray}
\noindent
where $|\psi_{i}\rangle\in {\cal H}$ may be expanded in the
harmonic oscillator basis as
$|\psi_{i}\rangle = \sum_{n=0}^{\infty}\,c_{n,i}|n\rangle$. The action of 
$P$ on the basis elements is easily determined to be
\begin{eqnarray}
 P|n\rangle \otimes |\One\rangle &=& \frac{1}{1+n}|n\rangle \otimes |\One\rangle +
 \frac{\sqrt{n}}{1+n}|n-1\rangle \otimes |\Two\rangle\nonumber\\
 P|n\rangle \otimes |\Two\rangle &=& \frac{\sqrt{1+n}}{2+n}|n+1\rangle \otimes 
 |\One\rangle + \frac{1+n}{2+n}|n\rangle \otimes |\Two\rangle
 \label{PActionBasis}
\end{eqnarray}
\noindent
so that in terms of the expansion coeffients $c_{n,1}$ and $c_{n,2}$ the eigenvalue
equation for $P$ becomes
\begin{eqnarray}
 &&\sum_{n=0}^{\infty}{\Bigl(}\frac{c_{n,1} - \lambda(n+1)c_{n,1} + \sqrt{n}c_{n-1,2}}
 {n+1}{\Bigr)}|n\rangle \otimes |\One\rangle\nonumber\\
 &+& \sum_{n=0}^{\infty}{\Bigl(}\frac{c_{n+1,1}\sqrt{n+1} + (n+1)c_{n,2} 
 - \lambda(n+2)c_{n,2}}
 {n+2}{\Bigr)}|n\rangle \otimes |\Two\rangle = 0\,.
 \label{E-Val-Eqn}
\end{eqnarray}
\noindent
Since $P$ is a projection operator, $\lambda = 0$ or $1$. Choosing first $\lambda = 1$ reduces (\ref{E-Val-Eqn}) to
\begin{eqnarray}
 c_{n,2} = \sqrt{n+1}c_{n+1,1}
\end{eqnarray} 
\noindent
which fixes completely the $c_{n,2}$ coefficients in terms of the $c_{n,1}$'s and gives 
\begin{eqnarray}
  |\psi_{1}\rangle &=& \sum_{n=0}^{\infty}\, c_{n,1}|n\rangle\nonumber\\
  |\psi_{2}\rangle &=&\sum_{n=0}^{\infty}\, \sqrt{n+1}c_{n+1,1}|n\rangle
\end{eqnarray}
\noindent
In an orthonormal eigenbasis $\{|\chi_{1}\rangle, |\chi_{2}\rangle\}$ the diagonal representation of  the $2\times2$ matrix $P$ is 
\begin{eqnarray}
 P =|\chi_{1}\rangle \langle \chi_{1}|\otimes|\One\rangle\langle\One | +
    |\chi_{2}\rangle \langle \chi_{2}|\otimes|\Two\rangle\langle\Two |
 \label{DiagonalP}
\end{eqnarray}
It remains only to fix the $c_{n,1}$ coefficients. This may be done by noting that $P$ is a solution of the $\CPn$ BPS equations only if $(1 - |\chi_{i}\rangle\langle\chi_{i}|)\widehat{a}|\chi_{i}\rangle\langle\chi_{i}| = 0$ {\it i.e.,} if $|\chi_{i}\rangle$ is an eigenstate of $\widehat{a}$. Written in terms of the expansion coefficients, this condition reads
\begin{eqnarray}
 c_{n,1} = \frac{\chi_{1}^{n}}{\sqrt{n!}}c_{0,1}
 \label{CoherentStateCoeffs}
\end{eqnarray}
\noindent
where $\chi_{1}$ is the eigenvalue corresponding to $|\chi_{1}\rangle$. This is, of course, expected of a coherent state in a harmonic oscillator basis. It may also quite easily be established that the $\lambda = 0$ case is trivial, yeilding $|\chi_{1}\rangle = |\chi_{2}\rangle = 0$. A straightforward application of Gram-Schmidt orthonormalization finds
\begin{eqnarray}
 |\chi_{1}\rangle = \sum_{n=0}^{\infty}\,\frac{c_{n,1}}{\sqrt{\sum_{m\geq 0}|c_{m,1}|^{2}}}|n\rangle\,,\qquad |\chi_{2}\rangle = 0
\end{eqnarray}
\noindent
leaving only the first term in (\ref{DiagonalP}). Without loss of generality, the residual degree of freedom in (\ref{CoherentStateCoeffs}) may be fixed by choosing $c_{0,1} = 1$. The eigenvalues $\chi_{1}$ are then interpreted as the complex location moduli of the solitons. For example, the simplest choice of $\chi_{1} = 0$ produces a degree $1$ soliton localized at the origin,
\begin{eqnarray}
 P=\left( \begin{array}{cc}
  |0\rangle\langle 0| &
  0\\
  0 &
  0
  \end{array} \right)
  \label{CP1-1-solitonAtOrig}
\end{eqnarray}
\noindent
The end result then is that in a diagonal representation the $\CP$ $1$-soliton solution is nothing but a unit rank GMS soliton. These results are easily extended to show that the $k$-soliton solution of the $\CPn$ sigma model in a diagonalizing basis may be written in the form (\ref{DiagP}). The interpretation here is that any degree $k$ $\CPn$ soliton may be built up of appropriate rank GMS solitons. Note, however, that the diagonalization is non unitary - given a rank $k$ matrix of the form (\ref{DiagP}) it is not possible in general to associate to it a unique (non-diagonal) Hermitian projection matrix that is also a solution of the sigma model equations. The set of solutions to the sigma model equations that are of the form (\ref{DiagP}) is considerably larger that those formed by adapted commutative constructions. As such, it is not surprising that the solution space of the noncommutative $\CPn$ sigma model is much larger than the corresponding commutative theory. In particular, as shown in \cite{FurutaInami2}, certain quasi-stable configurations of GMS solitons and anti-solitons form non-BPS states of of the noncommutative $\CPn$ sigma model that have zero size in the vanishing $\theta$ limit. Such solutions cannot be realized as the diagonalization of any non-diagonal solution of the sigma model equations.

\section{Conclusions and Discussion}
\label{Conclusions}
\noindent
In trying to understand the connection (if any) between solitonic excitations of the noncommutative sigma model on $\CPn$ and $D$-brane configurations in string theory, we have reformulated the noncommutative $\CPn$ model of \cite{LeeLeeYang} in a way that makes manifest the similarities (and differences) with the GMS scalar field theory. In doing so it becomes evident that the BPS solitons of the sigma model are no more immune from problems in the definition of the toplogical charge than any of the higher codimension solitons of, say, four-dimensional noncommutative gauge theory \cite{Ishikawa}. In this case, the naive calculation of the topological charge is in fact incorrect and must be supplemented by the addition of a nonvanishing ``surface term'' of the form ${\rm Tr}_{\cal H}{\rm tr}[\widehat{a},\cdot]$. Such terms vanish for GMS solitons and are consequently dropped in that case.\\

\noindent
We have also extended the SDZ construction for non-BPS solitons of the $\CPn$ model from known holomorphic (BPS) lumps and constructed explicit solutions for the case of $\CP$ and $\C\Pk^{2}$ . Unlike the commutative case though, the noncommutative SDZ construction does not yield the most general solitonic solutions of the sigma model equations. This incompleteness is due largely to the emergence of a new length scale in the problem as set by the noncommutativity parameter $\theta$. Evidently, as shown in \cite{FurutaInami2}, the solution space of the noncommutative sigma model is significantly larger than the corresponding commutative one with the additional (non-BPB) solitons made up of quasi-stable bound states of GMS solitons. While this construction might seem independent of the standard one, we have shown that, at least for the case of BPS solitons, they arise in the diagonalization of the Hermitian projector associated to a given BPS soliton.\\

\noindent     
The (commutative) $\CPn$ model is well known to arise as the low energy (infinite coupling) limit of a gauged linear sigma model with Fayet-Illiopolous $D$-term \cite{Schroers}. One class of solitonic excitations of this model are the votex solutions of the first order BPS equations
\begin{eqnarray}
 F_{12} + e^{2}(\phi^{\dagger}\phi - 1) &=& 0\nonumber\\
 D_{\overline{z}}\phi &=& 0\nonumber\\
 \Phi = \int\,d^{2}x F_{12} &=& 2\pi k > 0
 \label{VortexEqns}
\end{eqnarray}
\noindent
In the $e^{2}\rightarrow\infty$ limit these vortex solitons decend to the usual $\CPn$ lumps (when $\phi$ is an $(n+1)$-component complex vector). It was recently demonstrated that the noncommutative version of the linear sigma model in question also exhibits vortex excitations that are solutions of the noncommutative vortex equations \cite{DongsuBak,Bak,Wadia,Tong}
\begin{eqnarray}
 1+ [C^{\dagger},C] &=& \gamma (\phi^{\dagger}\phi - 1)\nonumber\\
 \phi a + C \phi &=& 0\nonumber\\
 {\rm Tr}_{\cal H}{\Bigl(}1 - [C^{\dagger},C]{\Bigr)} &=& -k
\end{eqnarray}
\noindent
where $\gamma = \theta e^{2}$ is a dimensionless parameter and $C$ is effectively the (noncommutative) Abelian gauge field. The exact vortex solutions of \cite{DongsuBak,Wadia,Tong} manifest in the $\gamma\rightarrow\infty$ limit (in which the vortex equations become tractable). Usually this limit is taken by sending $\theta\rightarrow\infty$ but clearly may also arise in the infinite coupling limit; the vortex equations are insensitive to which. As such, it is not unreasonable to expect that the exact vortex solutions decend to the lump solutions of the noncommutative $\CPn$ sigma model. Moreover, the noncommutative Abelian Higgs model may be embedded in a $(5+1)$-dimensional, ${\cal N}=1$ supersymmetric theory so that the vortices of the former become BPS $3$-branes which preserve half of the supersymmetries. So, more than just another academic exercise, the study of the infinite coupling limit of the noncommutative Abelian Higgs model may provide valuable insight into a string theoretic interpretation of the noncommutative $\CPn$ lumps. These issues will be addressed in future work.\\

\noindent
Yet another intriguing avenue for a stringy interpretation of the BPS solitonic excitations of the $\CPn$ sigma model is that offered by the work of \cite{LP3}. Drawing on the (tree level) equivalence of $N=2$ open string theory and self-dual Yang-Mills theory in $(2+2)$-dimensions \cite{OV} it was argued that the effective field theory induced by open $N=2$ strings in a K\"ahler $B$-field background on the worldvolume of $n$ coincident $D2$-branes is a modified $U(n)$ sigma model. The latter was also shown to exhibit solitonic solutions which were elegantly constructed using a `dressing method' \cite{LP1,LP2}. From this perspective, a string theoretic interpretation already exists: an $m$-soliton solution to the noncommutative $\CPn$-sigma model should correspond to $m$ $D0$-branes inside $(n+1)$ coincident $D2$-branes. A positive identification of the solitonic excitations of the sigma model with the $D0-D2$ system would, however, require more than just a matching of the energies of the two systems; it remains to compute the fluctuation spectra around the respective configurations. This is certainly an exciting avenue and warrants further research\footnote{We thank O. Lechtenfeld for several useful comments in this regard.}.  
 
\acknowledgments
 J.M. acknowledges support by the Sainsbury-Lindbury Trust and a research
 associateship of the University of Cape Town. R.A is supported by a University of Cape Town research associateship. We would like to 
 thank Philip Candelas, George Ellis, Bogdan Florea, Robert de
 Mello Koch, Joao Roderigues, David Tong and especially Amanda Weltman for useful discussions and valuable insights. 
 J.M. is especially grateful to Philip Candelas, for his kind hospitality at 
 the Mathematical Institute at Oxford University and to ISCAP (Columbia University) for kind hospitality during the time this work was being completed.    

\end{document}